\documentclass[aps,prd,a4paper,10pt,twocolumn,floatfix,showpacs,nofootinbib]{revtex4-1}
\usepackage{amsmath,amsfonts,amssymb,mathrsfs}
\usepackage{graphicx}
\usepackage{subfigure}
\usepackage{enumerate}
\usepackage{color}
  
\def\be{\begin{equation}}
\def\ee{\end{equation}}

\def\beq{\begin{equation}}
\def\eneq{\end{equation}}
\def\bea{\begin{eqnarray}}
\def\enea{\end{eqnarray}}
\def\nn{\nonumber}

\def\bea{\begin{eqnarray}}
\def\eea{\end{eqnarray}}
\def\nn{\nonumber}

\begin{document}

%%%%%%%%%%%%%%%%%%%%%%%%%%%%%%%%%%%%%%%%%%%%%%
\title{On power-counting renormalizability of Ho\v rava gravity with detailed balance}
%%%%%%%%%%%%%%%%%%%%%%%%%%%%%%%%%%%%%%%%%%%%%%

\author{Daniele Vernieri}
\affiliation{UPMC-CNRS, UMR7095, Institut d'Astrophysique de Paris, GReCO, 98bis boulevard Arago, F-75014 Paris, France}

%%%%%%%%%%%%%%%%%%%%%%%%%%%%%%%%%%%%%%%%%%%%%%
\begin{abstract}
We consider the version of Ho\v rava gravity where ``detailed balance'' is consistently implemented, so as to limit the huge proliferation of couplings in the full theory and obtain healthy dynamics at low energy. Since a superpotential which is third-order in spatial derivatives is not sufficient to guarantee the power-counting renormalizability of the spin-0 graviton, one needs to go an order beyond in derivatives, building a superpotential up to fourth-order spatial derivatives. Here we perturb the action to quadratic order around flat space and show that the power-counting renormalizability of the spin-0 graviton is achieved only by setting to zero a specific coupling of the theory, while the spin-2 graviton is always power-counting renormalizable for any choice of the couplings. This result raises serious doubts about the use of detailed balance.  
\end{abstract}
%%%%%%%%%%%%%%%%%%%%%%%%%%%%%%%%%%%%%%%%%%%%%%

\pacs{04.60.-m, 	% Quantum gravity
04.50.Kd,         %	Modified theories of gravity 
11.30.Cp          %	Lorentz and Poincare invariance
}

%%%%%%%%%%%%%%%%%%%%%%%%%%%%%%%%%%%%%%%%%%%%%%
\maketitle
%%%%%%%%%%%%%%%%%%%%%%%%%%%%%%%%%%%%%%%%%%%%%%

Ho\v rava gravity~\cite{Horava:2008ih,Horava:2009uw} has attracted a lot of attention since it was first proposed, as it encodes all the necessary ingredients to be both a renormalizable theory of gravity and a phenomenologically viable one (see Refs.~\cite{Sotiriou:2010wn,Mukohyama:2010xz,Visser:2011mf} for some reviews). 

The fundamental aim of the theory is to be an ultraviolet (UV) completion of general relativity, pursued by abandoning the local Lorentz invariance. It is based on the idea of modifying the graviton propagator by adding to the gra\-vi\-ta\-tio\-nal action higher-order spatial derivatives without adding higher-order time derivatives. In this way one can obtain a power-counting renormalizable theory~\cite{Visser:2009fg,Visser:2009ys}. Indeed, at the moment there is no definite evidence that the theory is fully quantum renormalizable (even if some evidence in this direction has been recently revealed in Ref.~\cite{D'Odorico:2014iha}), and the renormalizability is only supported by power-counting arguments.  

Since the theory treats space and time on different footing, it is naturally constructed in terms of a preferred foliation of spacetime, leading to violations of Lorentz symmetry at all scales. The theory is built using an ADM decomposition of spacetime,
\vspace{0.1cm}
\beq
ds^2=-N^2dt^2+ g_{ij} \left(dx^i+ N^i dt\right)\left(dx^j+ N^j dt\right),
\eneq
where $N$ is the lapse function, $N^i$ the shift vector, and $g_{ij}$ the induced three-dimensional metric on the spacelike hypersurfaces. 
  
The most general action of Ho\v rava gravity can be written as follows:
\beq
S=S_K-S_V,
\eneq
where 
\bea
S_K &=& \frac{2}{k^2}\int{dt d^3x\sqrt{g} N \left(K_{ij}K^{ij}-\lambda K^2\right)} \nn \\
&=& \frac{2}{k^2}\int dtd^3x\sqrt{g} N K_{ij}G^{ijkl}K_{kl},
\eea
is the kinetic term, which is quadratic in the time derivatives, $k$ is a coupling of suitable dimensions, $K_{ij}$ is the extrinsic curvature of the spacelike hypersurfaces,
\beq
K_{ij} = \frac{1}{2N} \left(\dot{g}_{ij}-\nabla_i N_j -\nabla_j N_i\right),
\eneq
$K = g^{ij} K_{ij}$ is its trace, $\nabla_i$ is the covariant derivative associated with $g_{ij}$, $\lambda$ is a dimensionless coupling, and $G^{ijkl}$ is the generalized DeWitt ``metric on the space of metrics", which is written in terms of the induced metric $g_{ij}$ as
\beq
G^{ijkl} = \frac{1}{2}\left(g^{ik}g^{jl}+g^{il}g^{jk}\right)-\lambda \, g^{ij}g^{kl}\,.
\eneq
The potential term is
\beq
S_V=\frac{k^2}{8}\int dt d^3x\sqrt{g} N \, V\left[g_{ij},N\right],
\eneq
and it includes all the operators built with the metric and the lapse compatibly with the invariance of the theory under foliation-preserving diffeomorphisms, {\it i.e.}, $t\rightarrow \tilde{t}(t)$, $x^i\rightarrow \tilde{x}^i(t, x^i)$. The most general action includes a very large number of operators $\sim \mathcal{O}(10^2)$ that are allowed by the symmetry. This makes the theory intractable and compromises predictability in the UV.
This is the reason why Ho\v rava in Ref.~\cite{Horava:2009uw} imposed some restrictions in order to limit the proliferation of couplings in the full theory. 
First, the {\it projectability} condition was assumed, which requires that the lapse function is just a function of time, {\it i.e.}, $N=N(t)$ (see Refs.~\cite{Sotiriou:2009gy,Sotiriou:2009bx} for the full implementation of projectability). 
Moreover, Ho\v rava also imposed an additional symmetry on the theory called {\it detailed balance}, which is inspired by condensed matter sy\-ste\-ms studied in the context of quantum criticality and nonequilibrium critical phenomena. The extension of this condition to the gravitational case sums up to the requirement that $V$ should be derivable from a superpotential $W$ as follows:
\beq
\label{Vdef}
V=E^{ij}{\cal G}_{ijkl}E^{kl}\,,
\eneq
where $E^{ij}$ is given in terms of the superpotential $W$ as
\beq
E^{ij}=\frac{1}{\sqrt{g}}\frac{\delta W}{\delta g_{ij}}\,,
\eneq
and ${\cal G}_{ijkl}$ is the inverse of the generalized DeWitt metric,
\beq
{\cal G}_{ijmn} G^{mnkl} = \frac{1}{2} \left(\delta_i^{\phantom{i}k} \delta_j^{\phantom{j}l} + \delta_i^{\phantom{i}l} \delta_j^{\phantom{j}k} \right),
\eneq
explicitly given by 
\beq
{\cal G}_{ijkl}=\frac{1}{2}\left(g_{ik}g_{jl}+g_{il}g_{jk}\right)+\frac{\lambda}{1-3\lambda} g_{ij}g_{kl}\,.
\eneq
The superpotential has to contain all of the possible terms which are invariant under foliation-preserving diffeomorphisms up to a given order in derivatives. The latter is dictated by the requirement that the theory is power-counting renormalizable, which happens when the number of spatial derivatives is $\geq 2d$, where $d+1$ indicates the dimensionality of the spacetime. In four dimensions, sixth-order spatial derivatives are then mi\-ni\-mal\-ly required in the action by power-counting arguments, which means that at least third-order spatial derivatives must be included in the superpotential. 

However, since the projectable version of the theory, with or without detailed balance, propagates a scalar degree of freedom which suffers instabilities and strong coupling at unacceptably low energies~\cite{Sotiriou:2009bx,Charmousis:2009tc,Blas:2009yd,Wang:2009yz,Koyama:2009hc}, in what follows we are not assuming projectability but only detailed balance. 

Our starting point will be the theory constructed from the following superpotential~\cite{Vernieri:2011aa,Vernieri:2012ms}: 
\beq \label{W}
W=\frac{1}{w^2}\int{\omega_3(\Gamma)}+\int{d^3x\sqrt{g}\left[\mu(R-2\Lambda_W\right)+\beta \, a_i a^i]}, 
\eneq 
where $\omega_3(\Gamma)$ is the gravitational Chern-Simons term, $a_i =\partial_i \mbox{ln} N$, and $w$, $\mu$, $\Lambda_W$, and $\beta$ are couplings of sui\-ta\-ble dimensions. This superpotential differs from the one used in the first proposal of the theory in Ref.~\cite{Horava:2009uw} for the operator controlled by the coupling $\beta$ which is not present there, since when projectability is implemented, $a_i=0$. The potential $V$ corresponding to the superpotential $W$ is then automatically given by using Eq.~\eqref{Vdef} (for the details, see Ref.~\cite{Vernieri:2011aa}). The resulting theory provides a well-behaved dynamics at low energy for both the spin-0 and the spin-2 graviton, choosing the coupling constants within suitable regions of the parameter space. In fact, it is just the presence of the extra coupling $\beta$ which makes healthy the infrared (IR) dynamics of the scalar degree of freedom (similarly to what happens in the most general theory, referred to as the healthy extension of Ho\v rava gravity~\cite{Blas:2009qj}). 

Nevertheless, it was shown in Ref.~\cite{Vernieri:2011aa} that the scalar does not satisfy a sixth-order dispersion relation but a fourth-order one, which is not sufficient to guarantee the power-counting renormalizability of the theory. In order to overcome this problem, it was conjectured that by adding fourth-order operators to the superpotential $W$, both sixth- and eighth-order terms would be generated in the potential $V$, thus rendering the theory power-counting renormalizable. However, such a conjecture needs to be checked with an actual calculation. This is the subject of the present paper in what follows. 

The fourth-order terms one can add to the superpotential $W$ can be written as
\bea \label{Wextra}
W_{extra} &=& \int d^3x\sqrt{g}\left[\gamma \, R^2 + \nu \, R^{ij}R_{ij} + \rho \, R \nabla^i a_i \right. \nn \\
&& + \, \chi \, R^{ij} a_i a_j + \tau \, R \, a_i a^i + \varsigma \left(a_i a^i \right)^2 + \sigma \left(\nabla^i a_i \right)^2  \nn \\ 
&& \left. + \, \theta \, a_i a_j \nabla^i a^j \right], 
\enea
which means that $8$ additional couplings have to be taken into account once the fourth-order operators are included in the superpotential. 

Moreover, the superpotential in Eq.~\eqref{W} gives rise to parity-violating terms in the action via the presence of the gravitational Chern-Simons term $\omega_3(\Gamma)$. Asking for invariance under parity transformations forces us to exclude such a term, and this is what we do in the fol\-lo\-wing. In Ref.~\cite{Vernieri:2011aa} it was not possible to exclude that term since it was necessary in order to guarantee the power-counting renormalizability of the spin-2 graviton, in absence of other higher-order operators that we are instead con\-si\-de\-ring here. So, in checking if the spin-0 graviton has the correct behavior in the UV, we should additionally verify that the spin-2 degree of freedom is behaving well, too. Notice that the IR behavior of the theory studied in Ref.~\cite{Vernieri:2011aa} is obviously unaffected by the addition of the aforementioned higher-order operators which will indeed introduce modifications to the dispersion relations from, at the most, the fourth-order in derivatives onwards. 

Let us now perturb the resulting action to quadratic order around a Minkowski background. We have
\bea
N=1+\alpha \,, \qquad  N_i=\partial_i y \,,  \qquad  g_{ij}=e^{2\zeta}\delta_{ij} + h_{ij}\,, \nn \\
\enea
where $\partial^i h_{ij} = \delta^{ij} h_{ij} = 0$, and we have used part of the available gauge freedom in order to eliminate the term $\partial_i\partial_j E$ in the most general scalar perturbation for $g_{ij}$ by setting $E=0$. Furthermore, the theory does not have vector excitations. 

At first-order for the extrinsic curvature $K_{ij}$ and its trace $K$, we obtain   
\bea
K_{ij}&=&\dot{\zeta}\delta_{ij}-\partial_i \partial_j y + \frac{1}{2}\dot{h}_{ij} \,,\\
K&=&3\dot{\zeta}-\partial^2 y \,,
\eea 
where $\partial^2 \equiv \delta_{ij}\partial^i\partial^j$, and for the Ricci tensor and the Ricci scalar of $g_{ij}$, we get 
\bea
R_{ij}&=&-\partial_i\partial_j\zeta-\delta_{ij}\partial^2\zeta-\frac{1}{2}\partial^2 h_{ij}\,,\\
R&=&-4\partial^2\zeta\,.
\eea
By using the full superpotential as given by the sum of Eqs. \eqref{W} and \eqref{Wextra} (with the exception of the $\omega_3(\Gamma)$ term that we are not considering), we can now write down the operators which contribute to the quadratic perturbative action, order by order in derivatives,
\bea
V^{(2)} =&& \frac{\mu^2\Lambda_W}{-1+3\lambda} R +\frac{\mu\Lambda_W\beta}{-1+3\lambda} a_i a^i\,,  \label{V2} 
\enea
\bea \label{V4}
V^{(4)} =&& \left(\mu^2 -\frac{\mu\Lambda_W\nu}{-1+3\lambda}\right)R_{ij}R^{ij} + \frac{\mu\Lambda_W\rho}{-1+3\lambda} a_i \nabla^i R  \nn \\
&+& \left[\frac{-4\mu\Lambda_W\gamma+\left(1-4\lambda\right)\mu^2}{4\left(-1+3\lambda\right)}\right] R^2  \nn \\
&-& \mu\Lambda_W \left[\frac{\chi+3\sigma}{-1+3\lambda}\right]\left(\nabla_i a^i\right)^2 \nn \\
&-& 2\mu\Lambda_W \left(\frac{\chi+\sigma}{-1+3\lambda}\right) a_i \nabla_j\nabla^i a^j  \nn \\
&-& \frac{\mu\Lambda_W\chi}{-1+3\lambda} \nabla_i a_j \nabla^j a^i - 2\mu\Lambda_W \left(\frac{\chi+4\tau}{-1+3\lambda}\right) a_i \nabla^2 a^i \nn \\
&-& 2\mu\Lambda_W \left(\frac{\chi+4\tau}{-1+3\lambda}\right) \nabla_i a_j \nabla^i a^j\,, 
\enea
\bea
V^{(6)} =&&\frac{\mu\left(\nu+8\gamma\lambda\right)}{2\left(-1+3\lambda\right)} R \nabla^2 R + \frac{2 \mu\lambda\rho}{-1+3\lambda} R \nabla^2 \nabla_i a^i  \nn \\
&-& 2\mu\left(\nu+2\gamma\right) R_{ij} \nabla^j\nabla^i R + 2 \mu\nu R_{ij} \nabla^2 R^{ij}  \nn \\ 
&-& 2\mu\rho \, R_{ij} \nabla^j\nabla^i\nabla_k a^k\,,  \label{V6}
\enea 
and 
\bea
V^{(8)} = &-&\left[\frac{16\gamma^2\left(1+\lambda\right)+16\gamma\nu + 3\nu^2}{4(-1+3\lambda)}\right](\nabla^2 R)^2  \nn \\
&+& (\nu+2\gamma)^2 \nabla_i\nabla_j R \nabla^i \nabla^j R  \nn \\
&+& 2\rho\left(\nu+2\gamma\right) \nabla_i\nabla_j R \nabla_k\nabla^i\nabla^j a^k  \nn \\
&-& 2\nu\left(\nu+2\gamma\right) \nabla_i\nabla_j R \nabla^2 R^{ij}  \nn \\
&-& 2\rho \left[\frac{\nu+2\gamma\left(1+\lambda\right)}{-1+3\lambda}\right] \nabla^2 R \nabla^2 \nabla_i a^i  \nn \\
&+& \rho^2 \, \nabla_i\nabla_j\nabla_k a^k \nabla_l\nabla^i\nabla^j a^l + \nu^2 \, \nabla^2 R_{ij} \nabla^2 R^{ij} \nn \\
&-& 2\nu\rho \, \nabla_i\nabla_j \nabla_k a^k \nabla^2 R^{ij} - \frac{\rho^2(1+\lambda)}{-1+3\lambda} \left(\nabla^2\nabla_i a^i\right)^2\,, \nn \\ \label{V8}
\enea
where $\nabla^2 \equiv g_{ij} \nabla^i\nabla^j$, and the number $^{(l)}$ just indicates the order in derivatives. There is also a bare cosmological constant term that we are omitting here as it is not relevant to the conclusions of this work. For a full discussion about the magnitude and sign of the bare cosmological constant, the reader can refer to Ref.~\cite{Vernieri:2011aa}.   

Let us first look at scalar perturbations. The variation of the action with respect to $y$ yields 
\beq
\partial^2 y=\frac{1-3\lambda}{1-\lambda}\dot{\zeta}\,.     
\eneq
Moreover, the variation with respect to $\alpha$ leads to
\begin{widetext}
\beq \label{alpha}
\alpha = \frac{-2 \mu^2 \Lambda_W + 2 \mu\rho \left[\Lambda_W + \left(\lambda-1\right)\partial^2\right] \partial^2 +2\rho\left(3\nu+8\gamma\right)\left(1-\lambda\right)\partial^6}{\mu \Lambda_W \beta + \mu \Lambda_W \sigma \partial^2 + 2\rho^2\left(1-\lambda\right)\partial^6}\zeta\,.
\eneq
\end{widetext}
In order to obtain the above equations, we have assumed suitable regular boundary conditions.
It follows that both $y$ and $\alpha$ are nondynamical auxiliary fields which can be integrated out in terms of $\zeta$. 

The perturbed potential term for the sixth- and eighth-order operators is, respectively, given by 
\bea
S_{V^{(6)}} =&& \frac{k^2}{8}\int dt d^3x \left[-\frac{4 \mu\left(-1+\lambda\right) \left(3\nu+8\gamma\right)}{-1+3 \lambda}\zeta \right. \nn \\
&+& \left. \frac{4\mu\rho\left(-1+\lambda\right)}{-1+3 \lambda}\alpha\right] \partial^6 \zeta
\enea
and
\bea
S_{V^{(8)}} =&& \frac{k^2}{8}\int dt d^3x \biggl\{\biggl[\frac{2\left(-1+\lambda\right)\left(3\nu+8\gamma\right)^2}{-1+3 \lambda}\zeta \biggr.\biggr. \nn \\
&-& \biggl. \frac{4\rho\left(-1+\lambda\right) \left(3\nu+8\gamma\right)}{-1+3\lambda}\alpha\biggr] \partial^8 \zeta \nn \\
&+& \biggl. \frac{2\rho^2 \left(-1+\lambda\right)}{-1+3\lambda} \alpha\, \partial^8 \alpha \biggr\}.  
\enea
Once $\alpha$ is integrated out by using Eq.~\eqref{alpha}, it is straightforward to show that the resulting dispersion relation for the scalar degree of freedom is at most fourth-order, since the higher-order contributions exactly cancel out. So the spin-0 graviton is not generically power-counting renormalizable, a quite unexpected result after having added the aforementioned higher-order operators to the superpotential $W$.  

Nevertheless, {\it if and only if $\rho =0$}, we get a dispersion relation where sixth- and eighth-order contributions are instead present, leading to
\beq
\omega_S^2 \sim \frac{1}{M_{pl}^4} \left(\frac{1-\lambda}{1-3\lambda}\right)^2 \left[2\mu\left(3\nu+8\gamma\right)p^6+\left(3\nu+8\gamma\right)^2 p^8 \right],
\eneq
where we have redefined the coupling constant $k$ in terms of the Planck mass as $k^2=4/M_{pl}^2$~\cite{Vernieri:2011aa}. Notice that the coefficient in front of $p^8$ is always positive and then it cannot lead to instabilities in the UV.

A separate discussion is needed on the spin-2 graviton. As previously stated, once parity invariance is imposed, one cannot still consider the Chern-Simons term in the original superpotential $W$. By con\-si\-de\-ring only the remaining operators in the superpotential $W$, one gets a dispersion relation for the spin-2 graviton which is not sixth-order anymore. However, ta\-king into account the extra terms in $W_{extra}$, it is straightforward to see that power-counting re\-nor\-ma\-li\-za\-bi\-li\-ty is generically preserved in the tensor sector. In fact, considering tensor perturbations we find that the ope\-ra\-tors $R_{ij} \nabla^2 R^{ij}$ in $V^{(6)}$ and $\nabla^2 R_{kl} \nabla^2 R^{kl}$ in $V^{(8)}$ generically yield non-trivial contributions, respectively, at sixth- and eighth-order, to the dispersion relation of the spin-2 graviton:  
\beq
\omega_T^2 \sim \frac{\nu}{M_{pl}^4}\left[-2\mu p^6 + \nu p^8\right].  \label{omegaT}
\eneq
So, we generically end up with a power-counting renormalizable theory for the spin-2 graviton. The latter is also classically stable at very high energies for any choice of the couplings since the coefficient in front of $p^8$ in Eq.~\eqref{omegaT} is always positive.  

In conclusion, we have found that the theory where detailed balance is consistently implemented, taking into account a superpotential with operators up to fourth-order in derivatives (which gives rise to a potential with operators up to eighth-order in derivatives), is not ge\-ne\-ri\-cal\-ly power-counting renormalizable as we would have expected. Indeed, the perturbations of the action to quadratic order around flat space lead to a dispersion relation for the scalar graviton which is neither of sixth- nor of eighth-order.  
Nevertheless, we have pointed out that it is still possible to achieve power-counting re\-nor\-ma\-li\-za\-bi\-li\-ty of the spin-0 graviton by setting to zero the coupling $\rho$ in front of the operator $R \nabla^i a_i$. 
On the contrary, the spin-2 graviton is always power-counting renormalizable for any choice of the couplings. 

This result makes us seriously question if detailed ba\-lan\-ce is really a viable condition one can impose in order to limit the huge proliferation of couplings in Ho\v rava gravity without compromising the power-counting renormalizability of the theory. In fact, the choice $\rho=0$, through which it is possible to restore the power-counting renormalizability, results to be a very strongly fine-tuned condition that we have to impose, especially if we want to keep the discussion fully general without the need of extra {\it ad hoc} assumptions. 

Moreover, in Refs.~\cite{Calcagni:2009ar,Calcagni:2009qw}, generalizations of detailed balance were considered by including matter fields which obey it. Also in such cases, serious concerns about the use of detailed balance were raised.

At any rate, in order to reduce the number of independent couplings in the full action of Ho\v rava gravity, one is in need of a principle or symmetry, and since the current suggestions of projectability and detailed balance do not seem optimal, further proposals in this direction are needed.

The author would like to thank Thomas P. Sotiriou for a critical reading of this manuscript. In the process of doing some calculations for this work the {\it xTras} package~\cite{Nutma:2013zea} developed by Teake Nutma for {\it Ma\-the\-ma\-ti\-ca} has been extensively used.
The research leading to these results has received funding from the European Research Council under the European Community's Seventh Framework Programme (FP7/2007-2013, Grant Agreement No. 307934).

\end{document}